\documentclass[pra,aps]{revtex4}
\usepackage{amsmath,amsfonts,amsthm,amssymb}

\newcommand\version{Dec. 21, 2004}

\newcommand\beq{\begin{equation}} 
\newcommand\eeq{\end{equation}}
\newcommand\Z{{\mathbb Z}} 

\newcommand\C{{\mathbb C}} 
\newcommand\Ss{{\mathbb S}}
\newcommand\R{{\mathbb R}} 
\newcommand\eps{\varepsilon}
\newcommand\half{\mbox{$\frac 12$}} 
\renewcommand\rho\varrho
 
\newcommand\kf{k_{\rm F}}
\newcommand\const{{\rm const.\, }} 
\newcommand\Tr{{\rm Tr\, }}
\newcommand{\n}{{\mathbf{\nabla}}}
\newcommand\BB{{\mathcal B}}
                                                 
\newtheorem{thm}{Theorem}
\newtheorem{lem}{Lemma}
\newtheorem{cor}{Corollary}
\theoremstyle{definition}
\newtheorem{rem}{Remark}                                          

\begin{document}

\title{Ground State Energy of the Low Density Fermi Gas}

\author{Elliott H. Lieb}
\affiliation{Department of Physics, Jadwin Hall, Princeton University,
P.O. Box 708, Princeton NJ 08544, USA}

\author{Robert Seiringer}
\affiliation{Department of Physics, Jadwin Hall, Princeton University,
P.O. Box 708, Princeton NJ 08544, USA}

\author{Jan Philip Solovej}
\affiliation{Department of Mathematics,
University of Copenhagen, Universitetsparken 5, DK-2100 Copenhagen,
Denmark}

\date{\version}

\begin{abstract}
  Recent developments in the physics of low density trapped gases make
  it worthwhile to verify old, well known results that, while
  plausible, were based on perturbation theory and assumptions about
  pseudopotentials.  We use and extend recently developed techniques
  to give a rigorous derivation of the asymptotic formula for the
  ground state energy of a dilute gas of $N$ fermions interacting with
  a short-range, positive potential of scattering length $a$. For spin
  $1/2$ fermions, this is
  $E \sim E^0 + (\hbar^2/2m) 2 \pi N \rho a$, where $E^0$ is the energy of the
  non-interacting system and $\rho$ is the density. A similar formula
  holds in 2D, with $\rho a$ replaced by $\rho /|\ln(\rho a^2)|$.
  Obviously this 2D energy is not the expectation value of a
  density-independent pseudopotential.
\end{abstract}


\maketitle

\section{Introduction}

The leading asymptotics for the ground state energy of a dilute gas of
fermions, interacting with a positive, short range pair potential, was
derived years ago by several approximate methods \cite{huang,lee,fetter}. 
Indeed, the leading correction beyond the ideal gas formula is no
different for fermions than for bosons, except for the fact that all
bosons interact with each other whereas the spin-up fermions
effectively interact only with the spin-down fermions and not with
each other. In this sense, the formula goes back to Lenz \cite{Lenz}
who derived the energy formula for bosons by assuming that each
particle interacts with $N-1$ fixed particles that are well spaced
from each other.

Thus, we expect that the ground state energy per unit volume, $e(\rho_\uparrow ,
\rho_\downarrow )$,  for $N_\uparrow$ spin-up
particles and $N_\downarrow$ spin-down particles of mass $m$ in a box
of volume $V$ (in the usual thermodynamic limit in which $V\to \infty$
and $\rho_\uparrow = N_\uparrow /V$ and $\rho_\downarrow =
N_\downarrow /V$ are fixed) is, asymptotically,
\beq\label{formula}
e(\rho_\uparrow , \rho_\downarrow ) = \frac{\hbar^2}{2m} 
\frac{3}{5}(6\pi^2)^{2/3}
\left(\rho_\uparrow^{5/3} + \rho_\downarrow^{5/3}\right) 
+ \frac{\hbar^2}{2m} 
 8\pi a \rho_\uparrow  \rho_\downarrow 
+\mathrm{higher\ order\ in\ } (\rho_\uparrow , \rho_\downarrow) \ ,
\eeq
where $a$ is the two-body ($s$-wave) scattering length of the pair
potential $v$. 
Under the assumption that the total density $ \rho \equiv
\rho_\uparrow + \rho_\downarrow $ is fixed, this formula indicates
that at low density the absolute ground state has spin zero, i.e., $
\rho_\uparrow = \rho_\downarrow =\rho /2 $.

The corresponding low density   formula for bosons contains  only  one
kind of density and is
\begin{equation}
e(\rho) =  \frac{\hbar^2}{2m} 4\pi a \rho^2
+\mathrm{higher\ order\ in\ } \rho \ .
\end{equation}
This asymptotic formula was proved rigorously in \cite{LY1998}.

It is customary, nowadays, to regard (\ref{formula}) as coming from a
pseudopotential $\frac{\hbar^2}{2m} 8\pi a \delta(x_i-x_j)$, and
that is certainly a useful shortcut to obtaining current results. But
this has to be justified mathematically, and that is the purpose of
this paper. Several issues of physical interest are involved, which
make it not totally obvious that the pseudopotential approach is
beyond need of justification.

\begin{itemize}
\item The availability of a good variational function $\Psi$ is
  important in theoretical physics, one that correctly displays the
  correlations of physical interest and whose energy $\langle \Psi| H|
  \Psi \rangle$ (which is necessarily an upper bound to the ground
  state energy) can be computed without resort to uncontrolled
  approximations. It should also give the correct energy to the
  desired accuracy.  A good example is the BCS function of
  superconductivity.  In the boson problem one would think of a
  Bijl-Dingle-Jastrow function $J =\prod_{i,j}g(x_i-x_j)$ but it has
  not been possible, as far as we know, to carry out the energy
  calculation without making assumptions.  The correlations are subtle
  (even if they are physically clear) and have to be treated
  carefully, and the required rigorous bosonic upper bound was finally
  found by Dyson \cite{dyson} but by using a non-bosonic variational
  function.  In the fermionic case considered here we use a function
  of the form $\Psi = S\cdot J$, where $S$ is a Slater determinant.
  While this $\Psi$ looks simple, the calculation of an upper bound of
  the required accuracy (\ref{formula}) occupies half of this paper,
  and one cannot say that this is a simple calculation.
  
\item The source of difficulties in the boson problem is the subtlety
  of the correlations, which constitute the entire energy. In the
  fermionic case, on the other hand, we are looking for a tiny
  correction to a dominant free-particle kinetic energy, but this
  contribution, especially for hard-core potentials, does not come
  from a small perturbation. It is not obvious that hard core
  collisions do not create energy changes by perturbing the Fermi
  surface.
  
\item The pseudopotential idea, while attractive, does have the
  drawback that it cannot be right in two dimensions (2D). The quantity $a
  \rho^2$ for 3D bosons is replaced by $\rho^2 / |\ln (\rho a^2)|$, as
  predicted by \cite{schick,hfm} and proved in \cite{LY2001}. (Note:
  The scattering length can be defined in 2D as well as 3D. See
  \cite{LY2001}.) Consequently, the pseudopotential will have to
  depend on $\rho$.  Thus, for fermions the energy to leading order in
  $\rho a^2$ ought to be 
\beq\label{formula2d} 
e(\rho_\uparrow ,
  \rho_\downarrow ) = \frac{\hbar^2}{2m} 2 \pi \left(\rho_\uparrow^{2}
    + \rho_\downarrow^{2}\right) + \frac{\hbar^2}{2m} \frac{8\pi
  }{|\ln ( \rho a^2)|} \rho_\uparrow \rho_\downarrow +\mathrm{higher\ 
    order\ in\ } (\rho_\uparrow , \rho_\downarrow) \ .  
\eeq 
We will prove formula (\ref{formula2d}) as well. Fermions in two
dimensional layers are physically interesting and it can be useful to
have (\ref{formula2d}) proved rigorously.

\item In addition to the pseudopotential approach there is also
the approach of summing diagrams \cite{baker}, which leads to many
terms beyond the two in (\ref{formula}). Nevertheless, it has to be
admitted that expansions, especially where hard-core potentials are
concerned, may have convergence or other difficulties. Effective field
theory methods have also been used successfully \cite{hammer}, but
with similar concerns. Therefore, rigorous confirmation is much to
be desired and we provide it here.
\end{itemize}

In the following we do everything in 3D until Section~\ref{2dsect},
where we explain the modifications necessary for 2D, some of which are
not trivial. This is done in order to make the essential ideas as
clear as possible. In a forthcoming paper \cite{RSfermiT}, the natural
generalization of (\ref{formula}) to positive temperature states will be
proved.


\section{Model and Main Results}

In units in which $\hbar^2/2m =1$ (which will be used henceforth), and
with $\Delta= \n^2$, the Hamiltonian is given by 
\beq H=\sum_{i=1}^N
-\Delta_i + \sum_{1\leq i<j\leq N} v(x_i-x_j), 
\eeq 
acting on anti-symmetric functions of $N$ space-spin variables, i.e.,
functions in $\bigwedge^N L^2(\Lambda;\C^q)$.  Here $q\geq 1$ denotes
the number of spin states.  The particles are confined to a bounded
region $\Lambda$, which we choose to be a cube of side length $L$ and
volume $V=L^3$ (or $L^2$ in 2D). We choose Dirichlet boundary
conditions for the Laplacian, i.e., $\Psi =0 $ when any $x_i$ is on
the boundary of the cube.

Since $H$ is spin-independent, we can specify the number of particles
of each spin, $N_1,\, N_2,\dots,\, N_q$ with $N=\sum_j N_j$. The wave
function $\Psi$ is then a function of the $N$ coordinates $x_1,\dots,
x_N$, without mention of spin at all, but with the requirement on
$\Psi $ that it be antisymmetric separately in the first $N_1$
variables, the second $N_2$ variables, etc. To avoid needless notation
we will give our proofs for $q=2$ but will state the main Theorems
\ref{T1} and \ref{T2} for general $q$.

The pair potential $v(x)$ is assumed to be positive, radial, and of
finite range $R_0$. It then has a finite and positive scattering
length $a$.  The scattering length can be defined as follows:
if $\varphi(x)$ is the unique solution (see \cite{LY2001} for a full 
discussion) of the
zero-energy scattering equation 
\beq\label{scatteq} 
-\Delta \varphi(x) +
\half v(x) \varphi(x) = 0 
\eeq 
subject to the boundary condition
$\lim_{|x|\to\infty} \varphi(x)=1$, then $a$ is given by $a=
\lim_{|x|\to\infty} |x| (1- \varphi(x))$.  Note that we do not assume
$v$ to be integrable; 
our results also apply to the case of a hard
core. Note also that for a pure hard-core interaction, the scattering
length is equal to the range.

There is no need (apart from simplicity) to assume that the potentials
$v(x)$ between different groups of particles are the same, thereby
allowing the Hamiltonian to be `spin-dependent'. Thus, we could take the
pair potential to be $v_{i,j}(x) $, (with $1\leq i,j\leq q$) 
between groups $i$ and $j$, with corresponding scattering lengths
$a_{i,j}$. In this way, the quantity $a\rho_i\rho_j$ in Theorem
\ref{T1} would be replaced by  $a_{i,j} \rho_i \rho_j$.
Our proof would still go through with obvious trivial changes. 

Our main result concerns the ground state energy $E_0(\{N_i\},L)$ of $H$,
in the thermodynamic limit
$L\to \infty$ with $\rho_i= N_i/L^3$ fixed. It is well known that
for systems with short range interactions the limit of 
the energy density, $ E_0(\{N_i\},L)/L_3$
exists and is independent of boundary conditions \cite{ruelle,robinson}.

\begin{thm}\label{T1}
Fix $\rho_i  =N_i/L^3$ for $1\leq i \leq q$ and $\rho = \sum_i \rho_i$,
and let $E_0(\{N_i\},L)$ denote the ground state energy of $H$
with the appropriate antisymmetry in each of the 
$N_i$ coordinate variables.
Then, for small $\rho$,
\beq
\lim_{L\to \infty} \frac{ 1}{L^3} E_0(\{N_i\}, L) = \frac 35 
\left( 6\pi^2 \right)^{2/3} \sum_{i=1}^q\rho_i^{5/3} 
+8\pi  a \sum_{1\leq i<j \leq q} \rho_i\, \rho_j
+a\rho^2 \eps(\rho),
\eeq
with  $-\const \left(a\rho^{1/3}\right)^{1/13}\leq \eps(\rho) 
\leq +\const \left(a\rho^{1/3}\right)^{2/9}$.
\end{thm}  

The constants in the bounds on $\eps(\rho)$ depend on the interaction
potential only through the
dimensionless ratio $R_0/a$. We could, in principle, display the
explicit dependence on $R_0/a$. By cutting off an infinite range
potential in an appropriate way, this would allow us to extend the
result (with different bounds on $\eps(\rho)$) to infinite-range
potentials with finite scattering length.

The analogous theorem in 2D is the following.
\begin{thm}\label{T2}
Fix $\rho_i  =N_i/L^2$ for $1\leq i \leq q$ and $\rho = \sum_i \rho_i$,
and let $E_0(\{N_i\},L)$ denote the ground state energy of $H$
with the appropriate antisymmetry in each of the 
$N_i$ coordinate variables.
Then, for small $\rho$,
\beq
\lim_{L\to \infty} \frac{ 1}{L^2} E_0(\{N_i\}, L) = 
2\pi  \sum_{i=1}^q\rho_i^{2} 
+\frac{8\pi}{|\ln( \rho a^2) |}  \sum_{1\leq i<j \leq q} \rho_i\, \rho_j
+\frac{\rho^2}{|\ln( \rho a^2)|} \eps(\rho),
\eeq
with  $-\const |\ln(a^2\rho)|^{-1/10}\leq \eps(\rho) 
\leq +\const |\ln(a^2\rho)|^{-1}\, \ln|\ln(a^2\rho)| $. 
\end{thm}

For simplicity we consider only $q=2$ henceforth, i.e., the spin
$\half$ case. The extension to general $q\geq 2$ is straightforward. We introduce the following convenient notation. 
For $N_1 + N_2 =N$, let $X=(x_1,\dots,x_{N_1})$ and
$Y=(y_1,\dots,y_{N_2})$ stand for the collection of spin-up and spin-down particle coordinates, respectively. The Hamiltonian can then be written as
\beq
H= -\Delta_X - \Delta_Y + v_{XX} + v_{YY} + v_{XY},
\eeq
with $\Delta_X = \nabla_X^2= \sum_{i=1}^{N_1} \n^2_{x_i}$,  $\Delta_Y=\nabla_Y^2=  \sum_{i=1}^{N_2} \n^2_{y_i}$,  $v_{XX}=\sum_{i<j} v(x_i-x_j)$, and $v_{XY} = 
\sum_{i,j} v(x_i-y_j)$. It acts on the Hilbert space of square-integrable
functions that are antisymmetric in the $X$ and in the $Y$ variables.

\section{Outline of Proof}

Before presenting the proof of Theorems~\ref{T1} and~\ref{T2} in full
detail, we give a short outline. We first concentrate on the
three-dimensional case. The proof is split into two parts, the upper
and lower bounds to the ground state energy. The upper bound, given in
Section~\ref{upsect}, uses the variational principle. The idea is to
construct a trial wave function that shows the features one would
expect the true ground state to have, but is at the same time
sufficiently simple to make it possible to compute a good upper bound
on the expectation value of the Hamiltonian.  For this latter purpose
we find it necessary to choose a function that confines the particles
to small boxes, separated from each other to avoid interaction
between different boxes. These boxes must not be chosen too small,
however, to ensure that the finite size effects are negligible
compared to the leading term in the interaction energy. Since this
latter energy is rather small, we are forced to have a large number of
particles in each of the boxes. This makes it impossible to control
the norm of our trial wave function, and hence we must carefully
take into account cancellations between the ratio of the expectation
value of the Hamiltonian and the norm of the trial wave function.

For the lower bound to the energy, given in Section~\ref{lowsect}, the
first essential step is to replace the \lq hard\rq\ interaction
potential $v(x)$ by a \lq soft\rq\ one, $W(x)$, at the expense of using up
some positive kinetic energy. This idea goes back to
Dyson~\cite{dyson}, who computed a lower bound for the ground state
energy of a hard-sphere Bose gas. Only the high-momentum part of the
kinetic energy is dispensable, however, since the low-momentum part is
needed to fill the Fermi sea.  In Lemma~\ref{dyson} below we prove
such a bound, using only momenta bigger than a certain cutoff. With
the soft potential $W(x)$ one can then hope to proceed with some sort of
{\it rigorous} perturbation theory to obtain a lower bound to the
energy. Indeed, we prove two {\it a priori} bounds, one on the
one-particle density matrix of the ground state, showing that it is
close to the projection onto the Fermi sea, and another one on the
number of particles whose distance to their nearest neighbor is small.
These bounds can be used to show that the ground state expectation of
$W(x)$ has the anticipated value.

The necessary modifications of our proofs for the corresponding result
in two dimensions, Theorem~\ref{T2}, are sketched in
Section~\ref{2dsect}.

\section{Upper Bound to the Ground State Energy}\label{upsect}

We start by collecting some properties of the solution to the 3D
zero-energy scattering equation (\ref{scatteq}). The proofs can be
found in the appendix of \cite{LY2001}.  The solution to (\ref{scatteq}), $\varphi(x)$, is
a radial function and satisfies
\begin{itemize}
\item [$\bullet$] $0\leq \varphi(x)\leq 1$, and hence $a>0$.
\item [$\bullet$]  $\varphi(x)$ is subharmonic on $\R^3$ (i.e., $\Delta \varphi(x) \geq 0$, see \cite{anal}), $\Delta \varphi(x)$ is a positive 
measure which is zero for $|x|>R_0$, and 
  $\int_{\R^3} \Delta \varphi(x)\, d^3\!x = 4\pi a $.
\item [$\bullet$]  $ \varphi(x) \geq 1 - a/|x|$, and $ \varphi(x) = 1 - a/|x|$
  for $|x|\geq R_0$.
\item [$\bullet$] $\int_{|x|\leq R} \left( |\n \varphi(x)|^2 +
\half v |\varphi(x)|^2 \right)\,d^3\!x = 4\pi a (1-a/R)$ for $R\geq R_0$. 
\end{itemize}

These properties will be useful both for the upper bound given in this
section and the lower bound given in the next.

For the upper bound, it will be convenient to localize the particles
into small boxes with Dirichlet boundary conditions. The number of particles in each box will be large
for small $\rho$, but finite and independent of the size of the large container $V$. 
Let the side length the small boxes be $\ell$. If we place these small boxes a distance $R_0$ from each other, then there will be no interaction between particles in different boxes. We then want to put $n = \rho_1 (\ell+R_0)^3$ spin-up particles into each box, and likewise $m=\rho_2 (\ell+R_0)^3$ spin-down particles. Since $\rho_i(\ell+R_0)^3$ need not be an integer, however, we will choose 
\begin{equation}\label{defnm}
n=\rho_1 (\ell+R_0)^3 +\eps_1 \quad {\rm and} \quad m=\rho_2 (\ell+R_0)^3 +\eps_2,
\end{equation}
with $0\leq \eps_1,\eps_2 < 1$ chosen such that $n$ and $m$ are
integers. We then really have too many particles, but this is legitimate for an upper bound, since the energy is certainly increasing with particle number.  
We thus have
\beq\label{upper}
\lim_{L\to\infty} \frac 1{L^3} E_0(N_1,N_2,L) \leq \frac{1}{(\ell+R_0)^3} E_0(n,m,\ell),
\eeq
with $n$ and $m$ given as in (\ref{defnm}). This bound holds for all choices of the box size $\ell$.

We will now derive an upper bound on the ground state energy of $n$ spin-up and $m$ spin-down particles in a cubic box of side length $\ell$, for general $n$, $m$ and $\ell$. 
We take as a trial state the function 
\beq\label{tri1}
\Psi(X,Y) = D_n(X) D_m(Y) G_n(X) G_m(Y) F(X,Y),
\eeq
where $D_n(X)$ denotes the Slater determinant of the first $n$
eigenfunctions of the Laplacian in a  cubic box of side length $\ell$, with
Dirichlet boundary conditions. (In the case of degeneracy, any choice will do.) Moreover,
\beq\label{tri2}
G_n(X)= \prod_{1\leq i<j\leq n} g(x_i-x_j),
\eeq
with $0\leq g(x)\leq 1$, having the property that $g(x)=0$ for $|x|\leq s$ and $g(x)=1$ for $|x|\geq 2s$, for some $s>2 R_0$ to be chosen later. We can assume that $|\nabla g(x)|\leq \const s^{-1}$ for some constant independent of $s$. Finally,
\beq\label{tri3}
F(X,Y) = \prod_{i=1}^n \prod_{j=1}^{m} f(x_i-y_j),
\eeq
with $f(x)=\varphi(x)/(1-a/R)$ for $|x|\leq R$ and $1$ otherwise. Here $\varphi(x)$ denotes the solution to the zero-energy scattering equation, and we assume that  $R>R_0$, which 
guarantees that $f$ is a continuous function. Moreover, we assume $2R\leq s$.  
By the variational principle,
\beq\label{vari}
E_0(n,m,\ell)\leq \frac {\langle\Psi|H|\Psi\rangle}{\langle\Psi|\Psi\rangle}.
\eeq

Since $\Psi$ vanishes whenever two particles of the same kind are closer together than the range of the interaction, we have
$$
\langle \Psi|H|\Psi\rangle =  \langle\Psi| -\Delta_X|\Psi \rangle + \langle\Psi| -\Delta_Y|\Psi \rangle  + \langle \Psi| v_{XY}|\Psi \rangle. 
$$
In evaluating the kinetic energy, we use partial integration and the fact that $D_n(X)$ is an eigenfunction of $-\Delta_X$. Let the corresponding eigenvalue (namely the sum of the lowest $n$ eigenvalues of the Dirichlet Laplacian) be denoted by $E^{\rm D}(n,\ell)$. Then
\begin{eqnarray*}
\langle \Psi|-\Delta_X|\Psi\rangle &=& E^{\rm D}(n,\ell) \langle\Psi|\Psi\rangle \\ && + \int D_n(X)^2 |\nabla_X G_n(X) F(X,Y)|^2 D_m(Y)^2 G_m(Y)^2 \,dX\,dY.
\end{eqnarray*}
Here we denoted $dX=\prod_{i=1}^n d^3\!x_i$ and $dY=\prod_{j=1}^m d^3\!y_j$ for short.
In the second term, we use the Schwarz inequality to deduce (for some $\eps>0$ to be chosen later)
\begin{eqnarray*}
|\nabla_X G_n(X) F(X,Y)|^2 &\leq& (1+\eps) |\nabla_X F(X,Y)|^2 G_n(X)^2 \\&& + \left( 1+\eps^{-1}\right) F(X,Y)^2 |\nabla_X G_n(X)|^2.
\end{eqnarray*}
Proceeding in the same way for the kinetic energy of the $Y$-particles, we thus get the upper bound
\beq\label{vari2}
\langle\Psi|H|\Psi\rangle \leq {\rm I} + (1+\eps) {\rm II} + \left(1+\eps^{-1}\right) {\rm III},
\eeq
with
\beq
{\rm I}=  \big[E^{\rm D}(n,\ell) + E^{\rm D}(m,\ell)\big] \langle\Psi|\Psi\rangle,
\eeq
\begin{equation}\label{defII}
{\rm II} = \int \left[ |\nabla_X F(X,Y)|^2  + |\nabla_Y F(X,Y)|^2  +  v_{XY} F(X,Y)^2\right] D_n(X)^2 D_m(Y)^2 G_n(X)^2 G_m(Y)^2\,dX\,dY,
\eeq
and
\beq\label{defIII}
{\rm III} = \int  \left[ |\nabla_X G_n(X)|^2 G_m(Y)^2 + |\nabla_Y G_m(Y)|^2 G_n(X)^2\right] F(X,Y)^2 D_n(X)^2 D_m(Y)^2\,dX\,dY.
\eeq
The positivity of $v_{XY}$ has been used here. We shall now bound these three terms, when divided by $\langle\Psi|\Psi\rangle$, separately. 

We start with ${\rm I}$. We may consider the sum of the $n$ lowest Dirichlet eigenvalues as a Riemann sum for the integral
$$
\left(\ell/\pi\right)^3 \int\limits_{|p|\leq \kf, \atop p_1,p_2,p_3\geq 0} p^2\,  d^3\!p=  \frac 3 5 (6\pi^2)^{2/3}\frac{n^{5/3}}{\ell^2},
$$ 
where we denote the Fermi momentum by $\kf=(6\pi^2 n)^{1/3}/\ell$. It is then easy to see that
\beq\label{estiI}
E^{\rm D}(n,\ell)\leq \frac 3 5 (6\pi^2)^{2/3}\frac{n^{5/3}}{\ell^2}\left(1+\const n^{-1/3}\right), 
\eeq
and the exponent in the error term is actually optimal. We note that this bound shows that we must not choose $\ell$ too small in order to have an error term that is negligible compared with $a\rho$; more precisely, we need $n\sim \rho_1 \ell^3  \gg (a^3\rho)^{-1}$. This will be fulfilled with our choice of $\ell$ below. 

Next we derive an upper bound on ${\rm II}$. We are going to need the following lemma.

\begin{lem}\label{lemdet}
Let $D_n(X)$ denote a Slater determinant of $n$ linearly independent functions $\phi_\alpha(x)$. For a given function $h(x)$ of one variable, let $\Phi(X)$ be the function $\Phi(X)=D_n(X)\prod_{i=1}^n h(x_i)$, and let $M$ denote the $n\times n$ matrix
\beq\label{defM}
M_{\alpha\beta} = \int \phi_\alpha^*(x) \phi_\beta(x) |h(x)|^2\, d^3\!x.
\eeq
Then
\begin{itemize}
\item [(i)] The norm of $\Phi$ is given by $\langle\Phi|\Phi\rangle = \det M$.
\item [(ii)] For $1\leq k\leq n$, the $k$-particle densities of $\Phi$ are given by
$$
\binom {n}{k} \frac 1{\langle\Phi|\Phi\rangle} \int |\Phi(X)|^2 \, d^3\! x_{k+1}\cdots d^3\!x_n =
\frac 1{k!} \prod_{i=1}^k |h(x_i)|^2 \big(x_1\wedge \cdots \wedge x_k \big| M^{-1} \otimes \cdots \otimes M^{-1} \big|  x_1\wedge \cdots \wedge x_k \big),
$$
where $|x)$ denotes the $n$-dimensional vector with components $\phi_\alpha(x)$, $1\leq \alpha\leq n$, and $|x_1 \wedge \cdots \wedge x_k)$ stands for the Slater determinant $(k!)^{-1/2} \sum_\sigma (-1)^{\sigma} |x_{\sigma(1)})\otimes \cdots\otimes |x_{\sigma(k)})$, $\sigma$ denoting permutations. 
\item [(iii)] If $\Phi'_i(X) = \Phi(X) k(x_i)/h(x_i)$ for some function $k(x)$, then
\beq\label{imme}
\sum_{i=1}^n \langle \Phi_i'|\Phi_i'\rangle = \Big(\det M \Big)\,\Big( \Tr[K M^{-1}] \Big), 
\eeq
where $\Tr[\, \cdot \, ]$ denotes the trace, and $K$ is the $n\times n$ matrix
\beq\label{defK}
K_{\alpha\beta} = \int \phi_\alpha^*(x) \phi_\beta(x) |k(x)|^2 \,d^3\!x.
\eeq
\end{itemize} 
\end{lem}

The proof of this lemma is a straightforward exercise that we leave to the reader. Note that without loss of generality one can set $h(x)=1$ in proving the lemma. Item (iii) is an immediate consequence of items (i) and (ii), noting that the left side of (\ref{imme}), when divided by the norm of $\Phi(X)$, is the integral of the one-particle density of $\Phi(X)$ multiplied by $|k(x)/h(x)|^2$; we state it as a separate item for later use.

Using $G_n(X)\leq 1$, we infer from this lemma that, for any fixed $Y$, 
\begin{eqnarray}\nonumber
&&\int  G_n(X)^2 D_n(X)^2 \left[ |\nabla_X F(X,Y)|^2 + \half v_{XY} |F(X,Y)|^2\right]\,dX \\ \nonumber
&&\leq \int  D_n(X)^2 \left[ |\nabla_X F(X,Y)|^2 + \half v_{XY} |F(X,Y)|^2\right]\,dX \\ &&=  \Tr[K_Y M_Y^{-1}] \, \int  D_n(X)^2 |F(X,Y)|^2 \,dX. \label{eq13}
\end{eqnarray}
The $n\times n$ matrices $K_Y$ and $M_Y$ are given by (\ref{defM}) and (\ref{defK}), with $\phi_\alpha(x)$ being the lowest $n$ Dirichlet eigenfunctions of $-\Delta$, and with $h(x)=\prod_j f(x-y_j)$ and 
$$
|k(x)|^2 = \big |\nabla_x \mbox{$\prod_j$} f(x-y_j)\big |^2+ \half \mbox{$ \sum_j$} v(x-y_j)\mbox{$\prod_j$} f(x-y_j)^2,
$$ 
respectively. 
Since $K_Y$ is a positive definite matrix, we have the bound $\Tr K_Y M_Y^{-1}\leq \|M_Y^{-1}\|\Tr K_Y$, where $\|\, \cdot \, \|$ denotes the matrix norm (i.e., the largest eigenvalue for hermitian matrices). To calculate $\Tr K_Y$, and to bound $\|M_Y^{-1}\|$, we can assume that all the $y_j$'s are separated by at least a distance $s$, because the integrand of term ${\rm II}$ in (\ref{defII}) vanishes otherwise. 
Since $s\geq 2R$ by assumption, we have in this case 
\beq\label{holds}
|k(x)|^2=\big|\nabla_x \mbox{$\prod_j$} f(x-y_j)\big|^2+ \half \mbox{$\sum_j$} v(x-y_j)\mbox{$\prod_j$} f(x-y_j)^2 
= \sum_{j=1}^{n} \xi(x-y_j)
\eeq
with
\beq\label{defxi}
\xi(x)= |\nabla f(x)|^2 + \half v(x) f(x)^2.
\eeq
Hence, if $\rho^{\rm D}_n(x)$ denotes the one-particle density of $D_n(X)$, we have
\beq\label{kytr}
\Tr K_Y = \sum_{j=1}^n \rho^{\rm D}_n* \xi (y_j),
\eeq
where $*$ denotes convolution. 

To bound $\|M_Y^{-1}\|$, we use the following:

\begin{lem}\label{lemmm}
Assume that $|y_i-y_j|\geq s\geq 2R $ for all $i\neq j$. Then
\beq
\| 1-M_Y \| \leq  \const \left(  \frac {a R^2}{s^3} +  n^{2/3} \frac {s^2}{\ell^2}\right).
\eeq
\end{lem}

\begin{proof}
Let $q(x)= 1-\prod_j f(x-y_j)^2 \geq 0$. Then, for any $n$-dimensional vector $|b)$ with components $b_\alpha$, 
$$
\big(b\big|1-M_Y\big|b\big) = \int  q(x) \Big |\sum_\alpha b_\alpha 
\phi_\alpha(x)\Big|^2 \,d^3\!x.
$$
Hence, the question about the largest eigenvalue of $1-M_Y$
translates into the question of how large the average potential energy for the
potential $q(x)$ can be for functions such as $\sum_\alpha b_\alpha \phi_\alpha(x)$ whose kinetic energy is bounded above by $({\rm const.})\, n^{2/3} \ell^{-2}$, i.e., the Fermi energy for $n$ particles.

Let $\BB_j$ denote the ball of radius $s/2$ around $y_j$. Note that all these balls are non-overlapping by assumption. Also, since $s\geq 2R$, $q(x)=0$ if $x$ is outside all the balls. For a given function $\phi(x)$, let $\phi_j$ denote the average of $\phi(x)$ in the ball $\BB_j$. Moreover, let $\eta(x) = \phi(x) - \bar \phi_j$. By the Cauchy-Schwarz inequality $(a+b)^2\leq 2(a^2+b^2)$, we get the bound 
\beq
\int_{\BB_j} q(x) |\phi(x)|^2 \,d^3\!x \leq 2 \int_{\BB_j} q(x) |\eta(x)|^2 \,d^3\!x + 2 |\phi_j|^2 \int_{\BB_j} q(x) \, d^3\!x.
\eeq
Note that $|\phi_j|^2 \leq 6/(\pi s^3) \int _{\BB_j} |\phi(x)|^2 \, d^3\!x$, again by the Cauchy-Schwarz inequality. Moreover, since $s\geq R$,
$$
\int_{\BB_j} q(x) \,d^3\!x = \int_{\R^3}(1-f(x)^2)\,d^3\!x \leq   (4\pi/3) aR^2.
$$
To obtain the last inequality, we used the definition of $f(x)$ as
well as the fact that $\varphi(x)\geq \max\{1-a/|x|\, ,\, 0\}$, as explained in
the beginning of this section.

Note that $\eta(x)$ is a function whose average over the ball $\BB_j$ is zero. In other words, it is orthogonal to the constant function in $\BB_j$. Hence, using the fact that $q(x)\leq 1$ and Poincar{\'e}'s inequality \cite{anal},
$$
\int_{\BB_j} q(x) |\eta(x)|^2\,d^3\!x \leq \int_{\BB_j}  |\eta(x)|^2\,d^3\!x \leq \const s^2 \int_{\BB_j}  |\nabla\eta(x)|^2 \,d^3\!x.
$$
In this last expression we can replace $\eta(x)$ by $\phi(x)$, of course, since they only differ by a constant. 
Summing over all the balls $\BB_j$ (and using that $q(x)=0$ outside the balls), we thus obtain that, for any function $\phi(x)$,
$$
\int_{\R^3}  q(x) |\phi(x)|^2 \,d^3\!x \leq  \const \left[ \frac {aR^2}{s^3}\int_{\R^3}  |\phi(x)|^2 \,d^3\!x + s^2 \int_{\R^3}  |\nabla\phi(x)|^2 \,d^3\!x \right].
$$
In the case in question, the kinetic energy of $\phi(x)$ is bounded by $\const n^{2/3}\ell^{-2}$. This finishes the proof of the lemma.
\end{proof}

Since $0\leq M_Y\leq 1$ as a matrix, this lemma implies that 
\beq\label{myesti}
\|M_Y^{-1}\| = \frac {1}{1-\|1-M_Y\|}\leq A_n \equiv \frac 1{ 1- \const \left[ aR^2 /s^3+ n^{2/3}(s/\ell)^2\right]}, 
\eeq 
provided the denominator is positive.
By inserting (\ref{kytr}) and (\ref{myesti}) into (\ref{eq13}), we see that, for fixed $Y$ with $|y_i-y_j|\geq s$ for all $i\neq j$,
\begin{eqnarray}\nonumber
&&\int  G_n(X)^2 D_n(X)^2 \left[ |\nabla_X F(X,Y)|^2 + \half v_{XY} F(X,Y)^2\right] \,dX\\ && \leq A_n \sum_{j=1}^n \rho^{\rm D}_n* \xi (y_j) \int  D_n(X)^2 F(X,Y)^2\, dX. \label{eq14}
\end{eqnarray}
To be able later to compare this expression (\ref{eq14}) with $\langle \Psi|\Psi\rangle$, 
we want to put $G_n(X)^2$ back into the integrand. For this purpose we need 
the following lemma, which compares the integrals with and without the factor $G_n(X)^2$. 

\begin{lem}\label{lemg}
For any fixed $Y$, 
\begin{eqnarray}\nonumber
&& \int  D_n(X)^2 F(X,Y)^2 G_n(X)^2\,dX \\ && \geq \int  D_n(X)^2 F(X,Y)^2 \,dX \, \left(1- \const n^{8/3} \|M_Y^{-1}\|^2  (s/\ell)^{5}\right). \label{normg}
\end{eqnarray}
\end{lem}

\begin{proof}
Since $g(x)=1$ for $|x|\geq 2s$, we have 
\beq\label{seco}
G_n(X)^2 \geq 1 - \sum_{i<j}^n \theta(2s - |x_i-x_j|).
\eeq
Here $\theta$ denotes the Heaviside step function, i.e., $\theta(t)=1$ for $t\geq 0$ and $\theta(t)=0$ for $t<0$.
To evaluate the integral of the second term in (\ref{seco}), we need the
two-particle density of the state $D_n(X)F(X,Y)$ for each fixed $Y$. By Lemma~\ref{lemdet}
above, and the fact that $f(x)\leq 1$, this density, when appropriately
normalized, is bounded from above by $\|M_Y^{-1}\|^2 \rho^{{\rm
  D},(2)}_n(x,x')$, where $ \rho^{{\rm D},(2)}_n(x,x')$ denotes the two-particle density of the determinantal state $D_n(X)$. In particular, by explicit computation one finds that this latter density satisfies the bound
\beq\label{2pdd} 
\rho^{{\rm D},(2)}_n(x,x') \leq \const |x-x'|^2
(n/\ell^3)^{8/3}  
\eeq 
for some constant independent of $n$ and $\ell$.
Hence we arrive at (\ref{normg}).
\end{proof}

We note that it is the $n$-dependence in inequality (\ref{normg}) that forces us to choose the particle number to be small and makes it necessary to localize the particles into small boxes. We also emphasize the importance of the exponent $5$ in (\ref{normg}), which stems from the fact that the two-particle density vanishes as $|x-x'|^2$ for $x$ close to $x'$. Had we not taken this into account, we would get an error term of the order $n^2 (s/\ell)^3$ in (\ref{normg}). Note that necessarily $s>a$, and hence this error would be huge if $n\gg (a^3\rho)^{-1}$, which is demanded by (\ref{estiI}). This also explains why it is not possible to treat $F(X,Y)$ on the same footing as $G_n(X)$ and $G_m(Y)$, since the two-particle density only vanishes as $|x-x'|^2$ for particles with {\it equal} spin, unlike the situation for particles of unequal spin. 

Let
$$
B_n= \left(1- \const n^{8/3} A_n^2  (s/\ell)^{5}\right)^{-1},
$$
assuming that the term in parenthesis is positive. 
Applying Lemma~\ref{lemg} to inequality (\ref{eq14}), we arrive at 
\begin{eqnarray}\nonumber
&&\int  G_n(X)^2 D_n(X)^2 \left[ |\nabla_X F(X,Y)|^2 + \half v_{XY} F(X,Y)^2\right] D_m(Y) G_m(Y)\,dX\,dY  \\ && \leq A_n B_n  \sum_{j=1}^n \int  \rho^{\rm D}_n* \xi (y_j) D_m(Y)^2 D_n(X)^2 F(X,Y)^2 G_m(Y)^2 G_n(X)^2 \,dX\,dY. \label{eq15}
\end{eqnarray}
Now we cannot bound $\rho^{\rm D}_n*\xi(y)$ independently of $y$ by simply using the supremum of $\rho^{\rm D}_n(x)$, since this number will be strictly bigger than $n/\ell^3$, even in the thermodynamic limit. Instead, we repeat the above argument for the $Y$ integration. We use $|G_m(Y)|\leq 1$, the $Y$-analogues of Lemma~\ref{lemdet} and then Lemma~\ref{lemg} to put $G_m(Y)^2$ back in. Here, it is important to note that now the $x_i$'s are separated by at least a distance $s\geq 2R$. In this way we  obtain
\begin{eqnarray}\nonumber
&&\int  G_n(X)^2 D_n(X)^2 \left[ |\nabla_X F(X,Y)|^2 + \half v_{XY} F(X,Y)^2\right] D_m(Y) G_m(Y)\,dX\, dY  \\ && \leq  A_n B_n B_m  \int  D_m(Y)^2 D_n(X)^2 F(X,Y)^2 G_m(Y)^2 G_n(X)^2 \, \Tr \widehat K_X M_X^{-1} \,dX\,dY. \label{ggg}
\end{eqnarray}
The matrix $M_X$ is the same as before, with $Y$ replaced by $X$ (and $n$ replaced by $m$, of course), and $\widehat K_X$ is the $m\times m$ matrix
$$
(\widehat K_X)_{\alpha\beta} = \int  \phi_\alpha(y)^* \phi_\beta(y) \prod_{i}f(y-x_i)^2 \rho^{\rm D}_n*\xi(y)\,d^3\!y.
$$
Using $|f(x)|\leq 1$ and $\|M_X^{-1}\|\leq A_m$, which follows from Lemma~\ref{lemmm} and the fact that the $x_i$'s are separated at least by a distance $s$, we get the bound
\beq\label{last}
\Tr \widehat K_X M_X^{-1} \leq A_m \Tr \widehat K_X\leq A_m \int \rho^{\rm D}_n(x) \rho^{\rm D}_m(y) \xi(x-y) \,d^3\!x\, d^3\!y.
\eeq

We recall the definition of $f(x)$ and the properties of $\varphi(x)$ stated in the beginning of this section to calculate the integral $ \int \xi(y)\,d^3\!y =  4\pi a (1-a/R)^{-1}$. We then use this information to bound the last integral in (\ref{last}), by using Young's inequality \cite{anal}. Thus 
\beq\label{inte}
\Tr \widehat K_X M_X^{-1} \leq A_m  \left( \int \rho^{\rm D}_n(x)^2 \,d^3\!x\right)^{1/2}\left( \int \rho^{\rm D}_m(y)^2 \,d^3\!y\right)^{1/2}   4\pi a (1-a/R)^{-1}.
\eeq
For the square of $\rho^{\rm D}_n(x)$ we find 
\beq\label{inte2}
\int  \rho^{\rm D}_n(x)^2\,d^3\!x = \frac 1 {\ell^3} \sum_{p,q} \prod_{a=1}^3 \left ( 1 + \half \delta_{p_a,q_a}\right),
\eeq
where $p_a$ denotes the components of the wave vector $p$, and the sums are over the $n$ lowest eigenstates of the Dirichlet Laplacian, $(2/\ell)^{3/2}\prod_{a=1}^3 \sin(p_a x_a)$. From this explicit expression it is easy to see that 
\beq\label{inte3}
\int \rho^{\rm D}_n(x)^2 \,d^3\!x \leq \frac {n^2}{\ell^3} \left (1+ \const n^{-1/3}\right).
\eeq
The same holds with $n$ replaced by $m$. Eq. (\ref{ggg}) thus implies the upper bound
\begin{eqnarray}\nonumber
&&\int  G_n(X)^2 D_n(X)^2 \left[ |\nabla_X F(X,Y)|^2 + \half v_{XY} F(X,Y)^2\right] D_m(Y) G_m(Y)\,dX\,dY  \\ &&  \leq \langle\Psi|\Psi\rangle \frac{ 4\pi a n m}{\ell^3} A_n A_m B_n B_m  (1-a/R)^{-1} \left (1+ \const n^{-1/3} + \const m^{-1/3}\right). \label{estiIIa}
\end{eqnarray}
The same bound holds, of course, with $X$ and $Y$ interchanged. We therefore have the upper bound
\beq
{\rm II}  \leq \langle\Psi|\Psi\rangle \frac{ 8\pi a n m}{\ell^3} A_n A_m B_n B_m  (1-a/R)^{-1} \left (1+ \const n^{-1/3} + \const m^{-1/3}\right). \label{estiII}
\eeq

It remains to bound the term ${\rm III}$. Using $|g(x)|\leq 1$ we have that  
\begin{eqnarray}\nonumber
|\nabla_X G_n(X)|^2&\leq&  \sum_{i=1}^n \ \sum_{j,\, j\neq i}
|\nabla g(x_i-x_j)|^2  \\ && + \sum_{i=1}^n \ \sum_{j,\, j\neq i} \ \sum_{k,\, k\neq i,j}
 |\nabla g(x_i-x_j)| |\nabla g(x_i-x_k)| . \label{susu}
\end{eqnarray}
Now, by Lemma~\ref{lemdet}, the appropriately normalized $k$-particle
densities of $D_n(X)F(X,Y)$ are bounded above by $\|M_Y^{-1}\|^k
\rho^{{\rm D},(k)}_n$, where $ \rho^{{\rm D},(k)}_n$ denotes the
$k$-particle density of $D_n(X)$. In particular, $\rho^{{\rm
    D},(2)}_n$ is satisfies the bound (\ref{2pdd}), and $\rho^{{\rm D},(3)}_n$
satisfies
$$
\rho^{{\rm D},(3)}_n(x,x',x'')\leq \const (n/\ell^3)^3
$$
for some constant independent of $n$ and $\ell$. 
Using the fact that $\nabla g(x)$ is supported on the set $|x|\leq 2s$, together with $|\nabla g(x)|\leq \const s^{-1}$, we obtain from (\ref{susu}), for any fixed $Y$,
\begin{eqnarray}\nonumber
&& \int  D_n(X)^2 F(X,Y)^2 |\nabla_X G_n(X)|^2\,dX \\ &&\leq \const \frac {n^2}{\ell^3} s \left( \|M_Y^{-1}\|^{2} n^{2/3} (s/\ell)^{2} + \|M_Y^{-1}\|^3 n (s/\ell)^3 \right) \int D_n(X)^2 F(X,Y)^2\,dX. \label{sec}
\end{eqnarray}
Finally, to get a bound on ${\rm III}$, we proceed as above, using (\ref{myesti}) (and the fact that the $y_j$'s are separated by a distance $s$) and Lemma~\ref{lemg} to put $G_n(X)^2$ back into the integral. Note, however, that it is enough to bound $A_n$ and $B_n$ by constants in this term. Assuming that $n(s/\ell)^3$ is small, the second term in the parenthesis in (\ref{sec}) is negligible compared to the first term. The same bound applies to the case where $X$ and $Y$ are interchanged, and hence we obtain
\beq\label{estiIII}
{\rm III} \leq \langle\Psi|\Psi\rangle\, \const \big(n^{8/3} +m^{8/3}\big)\frac {s^3}{\ell^5} .
\eeq

Collecting all the error terms obtained in Eqs. (\ref{estiI}), (\ref{estiII}) and (\ref{estiIII}) and inserting them into (\ref{vari}) and (\ref{vari2}), we obtain 
\begin{eqnarray}\nonumber
 E_0(n,m,\ell) &\leq& \frac 3 5  (6\pi^2)^{2/3} \frac{n^{5/3}+m^{5/3}}{\ell^2}\left(1+C n^{-1/3}+C m^{-1/3}\right) \\ \nonumber &&+ 8\pi a\frac {nm}{\ell^3}\left( 1 + \eps+  C \left[ \frac {a R^2}{s^3}+ (n+m)^{2/3} (s/\ell)^2 +\frac aR +\frac 1{n^{1/3}}+\frac 1{m^{1/3}} + (n+m)^{8/3} (s/\ell)^5 \right]\right)\\  &&+ \frac {Cs}\eps \frac { (n+m)^2}{\ell^3} \left[ (n+m)^{2/3} (s/\ell)^2\right]  \label{ineq33}
\end{eqnarray}
for some constant $C>0$. In Ineq. (\ref{ineq33}) we have assumed smallness of  all the error terms, i.e., that the terms in square brackets are small. This condition will be fulfilled, at low density, with our choice of $R$, $s$, $n$, $m$ and $\ell$ below. 

The optimal choice of $\eps$ in (\ref{ineq33}) is given by $\eps^2 = \const (n+m)^{8/3} s^3/(\ell^2 a n m)$. Inserting this value for $\eps$ we infer from (\ref{ineq33})
\begin{eqnarray}\nonumber
 E_0(n,m,\ell) &\leq& \frac 3 5  (6\pi^2)^{2/3} \frac{n^{5/3}+m^{5/3}}{\ell^2}\left(1+C n^{-1/3}+C m^{-1/3}\right) \\ \nonumber &&+ 8\pi a\frac {nm}{\ell^3}\left( 1 +  C \left[ \frac {a R^2}{s^3}+ (n+m)^{2/3} (s/\ell)^2 +\frac aR +\frac 1{n^{1/3}}+\frac 1{m^{1/3}} + (n+m)^{8/3} (s/\ell)^5 \right]\right)\\  &&+ C(n+m)^{7/3} \frac {s^{3/2} a^{1/2}}{\ell^4} .  \label{ineq33s}
\end{eqnarray}
Eq. (\ref{ineq33s}) is our final bound on the energy $E_0(n,m,\ell)$. To apply this result in (\ref{upper}) we have to insert the values (\ref{defnm}) for $n$ and $m$. 
Recall that $|n - \rho_1 (\ell+R_0)^3|\leq 1 $ and $| m -   \rho_2 (\ell+R_0)^3|  \leq 1$. We are then still free to choose $R$, $s$ and $\ell$. We choose
$$
R=a \big(a\rho^{1/3}\big)^{-2/9}\ , \ s=2 R \ , \ 
 \ell = \rho^{-1/3} \big(a\rho^{1/3}\big)^{-11/9}.
$$
Inserting these values into (\ref{ineq33s}) we thus obtain, for small $\rho$, 
$$
\frac 1{\ell^3} E_0(n,m,\ell) \leq 
\frac 3 5  (6\pi^2)^{2/3} \big[\rho_1^{5/3}+ \rho_2^{5/3}\big]+ 8\pi a \rho_1 \rho_2
+ \const  a \rho^2  \big(a\rho^{1/3}\big)^{2/9}.
$$
In combination with Eq. (\ref{upper}), this finishes the proof of the upper bound. Note that the contribution to the error term that arises from the fact that $E_0(n,m,\ell)$ has to be divided by $(\ell+2R_0)^3$ and not $\ell^3$ in (\ref{upper}), is of the order $\rho^{5/3} R_0/\ell$ and, for this choice of $\ell$, is much smaller than $a\rho^2 (a\rho^{1/3})^{2/9}$ when $\rho$ is small.

\section{Lower Bound to the Ground State Energy}\label{lowsect}

\subsection{The Dyson Lemma}

We start with a generalization of a lemma
of Dyson \cite{dyson}, which bounds the hard potential $v(x)$ (which may or
may not contain a hard core) from below by a soft potential $U(x)$, at
the expense of using up some positive kinetic energy. In the following,
$\widehat f(k)$ denotes the Fourier transform of a function $f(x)$, i.e.,
$\widehat f (k) =(2\pi)^{-3/2}\int f(x) \exp (ik\cdot x) \,d^3\!x$.
Dyson's result was generalized in \cite{LY1998} and it is further
generalized here by separating low from high momentum. In our
application ``low'' will mean $\rho^{1/3}$ and ``high'' will mean
$1/a$. The analogous inequality for 2D is stated as Lemma~\ref{dyson2d} below. The proof of both lemmas is given in the appendix. 

\begin{lem} \label{dyson}
For $R>R_0$, let $\theta_R(x)$ denote the characteristic function 
of a ball 
of radius $R$ centered at the origin, i.e., $\theta_R (x) =1 $ if 
$|x| <R$ and $=0$ otherwise. Let $\chi(p)$ be a radial function, 
$0\leq \chi (p) \leq 1$, such that $h(x)\equiv \widehat{(1-\chi)}(x)$ 
is bounded and
integrable.
Let 
  \beq\label{deffr} 
f_R(x)= \sup_{|y|\leq R} | h(x-y) - h(x) |, 
\eeq
  and 
\beq\label{defwr} w_R(x)= \frac{2}{\pi^2} f_R(x) \int_{\R^3}
  f_R(y)\,d^3\!y.  
\eeq 
Then, for any positive, radial function $U(x)$,
  supported in the annulus $R_0\leq |x|\leq R$, with $\int_{\R^3}
  U(x)\,d^3\!x = 4\pi$, and for any $\eps>0$,
\beq\label{topr}
-\nabla \chi(p) \theta_R(x) \chi(p) \nabla + \half v(x) 
\geq (1-\eps) a U(x) - \frac 
a\eps w_R(x).
\eeq
Here, $\theta_R(x)$ is  multiplication operator in $x$-space, whereas 
$\chi(p)$ is a multiplication operator in momentum space. 
Thus, $\nabla \chi(p) \theta_R(x) \chi(p) \nabla$ is an operator
version of $\nabla^2$, which is cut off in both configuration and in momentum
spaces.
\end{lem}

The original Dyson lemma (as modified in \cite{LY1998}) has $\chi(p)\equiv 1$ and $w_R(x)\equiv 0$, i.e., there is no cutoff. The cutoff $\chi(p)$ in (\ref{topr}) essentially says that only the high momentum part of $\nabla$ is needed to give a good account of the scattering of two particles. The (relatively) low momentum part of $\nabla$ is not used in (\ref{topr}) and is thereby saved for later use to give a good estimate of the 
part of the fermion kinetic energy needed to fill the Fermi sea.  The price we pay for this luxury is the error term $a w_R(x)/\eps$, which does not appear in Dyson's lemma. 

Note that, by construction, either $w_R(x)$ is bounded and integrable or else
$w_R(x)=\infty$ for all $x$.  If $\chi(p)\equiv 1$, then $w_R(x)\equiv
0$, and hence we can set $\eps=0$ to recover the Dyson Lemma in
\cite{LY1998}, which says that for any $\phi(x)$, \ $\int_{|x|<R}(|\nabla
\phi (x)|^2 +[\frac 12 v(x) -a U(x)] |\phi(x)|^2) \,d^3\!x
\geq 0$. 

\begin{cor}\label{wcor}
If $y_1,\dots,y_N$ denote  $N$  points in $\R^3$, with $|y_i-y_j|\geq 2R$ 
for all $i\neq j$, then, as an operator on functions of $x$,
\beq
-\nabla \chi(p)^2 \nabla + \half \sum_{i=1}^N v(x-y_i) \geq  
\sum_{i=1}^N 
\left( (1-\eps) a U(x-y_i) - \frac a\eps w_R(x-y_i) \right).
\eeq
\end{cor}

\begin{proof}
This follows immediately from the previous lemma, using translation
  invariance and the fact that $\sum_i \theta_R(x-y_i)\leq 1$ since
  all the balls are non-overlapping, by assumption.
\end{proof}

To apply this corollary, let $l(p)$ be a smooth, radial, positive
function with $l(p)=0$ for $|p|\leq 1$, $l(p)=1$ for $|p|\geq 2$, and
$0\leq l(p)\leq 1$ in-between. For some $s>0$ let 
\beq\label{defchi}
\chi_s(p)=l(s p).  
\eeq 
Note that with this choice of $\chi_s(p)$ the corresponding
$h(x)=\widehat{1-\chi_s}(x)$ is a smooth function of rapid decay and hence,
by simple scaling, the corresponding potential $w_R(x)$ satisfies (for
$R\leq \const s$) 
\beq\label{intw} 
 |w_R(x)| \leq \const \frac{R^2}{s^5} \ \quad
{\rm and\quad } \int |w_R(x)|\,d^3\!x \leq \const \frac {R^2}{s^2} 
\eeq 
for some
constants depending only on $l$. Moreover, if $|y_i-y_j|\geq 2R$ for
all $i\neq j$, then 
\beq\label{normw} 
\sum_{i=1}^N w_R(x-y_i) \leq \const \frac 1{R s^2} 
\eeq 
independent of $x$ and $N$. 
Later we are going to choose $R\ll s\ll \rho^{-1/3}$ (cf. Eq.~(\ref{chos})).

\subsection{A priori bounds}\label{apss}

For $N_1+N_2=N$, let $\Psi_N(X,Y)$ be a sequence of normalized wave functions, antisymmetric both in the $X$ and in the $Y$ variables. We assume that $N_1/L^3\to \rho_1$ and $N_2/L^3\to \rho_2$ as $N\to\infty$, with $\rho=\rho_1+\rho_2$. Let $\gamma_1$ and $\gamma_2$ denote the reduced
one-particle density matrices of $\Psi_N(X,Y)$ for the $X$- and
$Y$-particles, respectively, with $\Tr \gamma_1=N_1$ and
$\Tr\gamma_2=N_2$. Moreover, let $P_M$ denote the following spectral
projection of the Laplacian with {\it periodic} boundary conditions on
$\Lambda$, given by the integral kernel 
\beq\label{perker} 
P_M(x,x')=
\frac 1{L^3} \sum_{p\in (2\pi/L) \Z^3 \atop |p|\leq (6\pi^2
  M/L^3)^{1/3}} e^{i p\cdot (x-x')} 
\eeq 
for $x,x'\in\Lambda$. Note that, by
scaling, $\Tr [P_M]$ does not depend on $L$, and 
\beq\label{trpm}
\lim_{m\to\infty} \frac 1 M \Tr[P_M] = 1.  
\eeq

In Lemmas~\ref{lemtr} and~\ref{lemir} we derive some bounds for sequences of wave functions satisfying certain energy bounds. These lemmas apply, in particular, to the true ground state -- as shown in the previous section. 
We call these bounds {\it a priori} bounds. 

\begin{lem}\label{lemtr}
Assume that, in the thermodynamic limit ($N\to\infty$,
$L\to\infty$ with $\rho_i=N_i/L^3$ fixed), there is a sequence of states $\Psi_N(X,Y)$ such that 
\beq\label{assu}
\limsup_{L\to \infty} \frac 1{L^3} \langle \Psi_N| H| \Psi_N\rangle \leq \frac 35 \left( 6\pi^2 \right)^{2/3} \left[ \rho_1^{5/3}+\rho_2^{5/3}\right] + C  a \rho^2
\eeq
for some $C>0$ independent of $\rho$. Then, for $i=1,2$ (and with $\gamma_1$, $\gamma_2$ being the one-body density matrices), 
\beq\label{cond}
\limsup_{L\to\infty} \frac 1 {L^3} \Tr[ \gamma_i(1-P_{N_i})] \leq \const 
\rho \sqrt { a \rho^{1/3} }. 
\eeq
\end{lem}

\begin{proof}
  We immediately have the trivial lower bound for non-interacting fermions
\beq\label{refi} 
\liminf_{L\to
    \infty} \frac 1{L^3} \langle \Psi_N| -\Delta_X-\Delta_Y|
  \Psi_N\rangle \geq  
\frac 35 \left( 6\pi^2 \right)^{2/3}
  \left[\rho_1 ^{5/3}+\rho_2
    ^{5/3}\right].
\eeq  
To prove Ineq.~(\ref{cond}), however, we need the following refinement of
(\ref{refi}), which is proved in \cite[Eq.~(4.13)]{GS}:
\begin{equation}
\liminf_{L\to \infty} \frac 1{L^3} \langle \Psi_N|
-\Delta_X-\Delta_Y| \Psi_N\rangle \geq   \frac 35 \left(
6\pi^2 \right)^{2/3} 
\limsup_{N\to\infty} \left[ \rho_1 ^{5/3}(1+\const
\zeta_1^2)+\rho_2 ^{5/3}(1+\const \zeta_2^2) \right],
\end{equation}
where $\zeta_i = N_i^{-1} \Tr[ \gamma_i(1-P_{N_i})]$ for $i=1,2$. Using (\ref{assu}) as well as the fact that the interaction potential is assumed to be positive, we immediately obtain (\ref{cond}).
\end{proof}

Our second {\it a priori} bound concerns the nearest neighbor
distance among particles. For given points $y_1,\dots,y_{N_2}$ in $\R^3$, let
$I_R(y_1,\dots,y_{N_2})$ be the number of $y_i$'s with the property that
the distance to the nearest neighbor among the other $y_i$'s is less
than $2R$.

\begin{lem}\label{lemir}
Assume that there exists a $C>0$, independent of $\rho=N/L^3$, such that
\beq
 \frac 1N \langle \Psi_N| H| \Psi_N\rangle \leq C \rho^{2/3}.
\eeq 
Then 
\beq\label{ineqI}
\langle \Psi_N| I_R(y_1,\dots,y_{N_2}) |
\Psi_N\rangle \leq \const N (R^3\rho)^{2/3}.
\eeq
\end{lem}

\begin{proof}
With $\delta_i$ denoting the distance to the nearest neighbor, we have
$$
I_R(y_1,\dots,y_{N_2}) \leq  (2R)^2 \sum_{i=1}^{N_2} \frac 1{\delta_i^2}.
$$
The result now follows from the operator inequality
\beq\label{lyineq} 
\sum_{i=1}^{N_2} \frac 1{\delta_i^2} \leq \const
\sum_{i=1}^{N_2} -\Delta_i 
\eeq 
which holds on anti-symmetric wave
functions of $N_2$ variables, and is proved in \cite[Thm.~5]{LYau}.
Note that we again used the fact that the interaction potential is
positive, and hence the kinetic energy is bounded above by the total
energy.
\end{proof}

\subsection{Putting it together}\label{ssput}

For a lower bound, we can neglect the interaction among particles of
the same kind. That is, we use 
\beq\label{hamxy} 
H\geq \left(
  -\Delta_X + \frac12 v_{XY} \right) + \left( -\Delta_Y + \frac 12
  v_{XY} \right).  
\eeq 
We are going to bound both terms separately
using the {\it a priori} bounds of the previous section.  In the following, we
are going to treat only the first term, the lower bound on the second term
can be obtained in the same way by exchanging $X$ and $Y$.

First, we decompose the Laplacian into a high and a low momentum part, as follows:
$$
\Delta = \nabla \Gamma(p) \nabla + \nabla \left(1-\Gamma(p)\right)\nabla. 
$$
For $\rho=N/L^3$, let $\kf= (6\pi^2\rho)^{1/3}$ be the Fermi momentum (for spinless fermions), and let 
$$
\Gamma(p)= \max\left\{ 1 - \frac {\kf^2}{|p|^2}\, , \, 0\right\}.
$$
We claim that  
\beq\label{lyclaim}
\sum_{i=1}^{N_1} -\nabla_{i} \left(1-\Gamma(p_i)\right)\nabla_i \geq \frac 35 \left(6\pi^2\right)^{2/3} \frac{N_1^{5/3}}{L^2}.
\eeq
To show this, we use the argument in \cite{liyau}. Let $\phi_i(x)$, $i=1,\dots,N_1$, be any set of orthonormal functions with support in the cube $\Lambda$ of side length $L$. We want a lower bound to the expression
\beq\label{ly}
\sum_{i=1}^{N_1} \int_{\R^3}  |p|^2 \left(1-\Gamma(p)\right) |\widehat \phi_i(p)|^2 \,d^3\!p .
\eeq
Note that $\widehat\phi_i(p)=(2\pi)^{-3/2} \langle e^{ip\cdot x}|\phi_i\rangle$, with $\langle\, \cdot \, | \, \cdot\,\rangle$ denoting the inner product for functions on the cube $\Lambda$. 
Since the $\phi_i(x)$ are orthonormal, we have that 
$$
\sum_{i=1}^{N_1} |\widehat \phi_i(p)|^2 \leq (2\pi)^{-3} \langle e^{ip\cdot x}|e^{ip\cdot x}\rangle =(2\pi)^{-3} L^3.
$$
Hence (\ref{ly}) is bounded below by the infimum of $\int  p^2 \left(1-\Gamma(p)\right)\xi(p)\,d^3\!p$ over all $0\leq \xi(p)\leq (2\pi)^{-3} L^3$ with $\int \xi(p)\,d^3\!p =N_1$. Since $|p|^2(1-\Gamma(p))$ is a monotone increasing function of $|p|$, the infimum is attained  by $\xi(p)=\theta ( (6\pi^2 N_1/L^3)^{1/3}-|p|)$, with $\theta$ denoting the Heaviside step function. Now $\Gamma(p)=0$ for $|p|\leq (6\pi^2 N_1/L^3)^{1/3} \leq (6\pi^2\rho)^{1/3}$, and thus we arrive at (\ref{lyclaim}).

For the high-momentum part, we use that
$$
\Gamma(p) \geq \left( 1- s^2 \kf^2\right) \chi_s(p)^2
$$
for any $s\leq 1/\kf$, with $\chi_s(p)$ defined in (\ref{defchi}). 
Hence, we can use Corollary~\ref{wcor} to get a lower bound on this
term. In order to be able to apply this corollary, however, we have to
make sure that the $y_j$'s are separated at least a distance $2R$. Let
$\widetilde Y \subset Y$ be the set of $y_j$'s whose distance to the
nearest neighbor is at least $2R$. Note that, by definition,
$|\widetilde Y|= N_2-I_R(Y)$. We are going to neglect the interaction
with $y_j$'s that are not in the set $\widetilde Y$, which can only
lower the energy. Hence we obtain, for a given configuration of $Y$, 
$$
\sum_{i=1}^{N_1}- \nabla_{i} \Gamma(p_i)\nabla_i + \frac 12 \sum_{i,j} v(x_i-y_j) \geq 
(1-s^2\kf^2) \sum_{i=1}^{N_1} W_{Y}(x_i),
$$
with 
\beq\label{defwy}
W_Y(x) = \sum_{\{ j\, : \, y_j \in \widetilde Y\}} \left( (1-\eps) a U(x-y_j) - \frac a\eps w_R(x-y_j) \right),
\eeq
depending on $\eps$, $a$, $R$ and $s$.  

We are still free to choose $U(x)$. A convenient choice is 
$$
U(x) = \left\{ \begin{array}{ll}  3\left(R^3-R_0^3\right)^{-1} & {\rm for\ } R_0\leq |x|\leq R \\ 0 & {\rm otherwise}, \end{array} \right.
$$
but any other choice such that $|U(x)|\leq \const R^{-3}$ for $R\gg R_0$ will do for our purpose. 

Now let $\Psi_N(X,Y)$ be a normalized fermionic wave function. We can express the expectation value of $\sum_i W_Y(x_i)$ as
\beq
\left\langle \Psi_N\left| \sum_{i=1}^{N_1} W_Y(x_i) \right| \Psi_N\right\rangle = \int n_Y \Tr[\gamma_Y W_Y] \,dY ,
\eeq
where 
\beq
n_Y = \int |\Psi_N(X,Y)|^2 \,dX
\eeq
and $\gamma_Y$ denotes the one-particle density matrix of $\Psi_N(X,Y)$ for {\it fixed} $Y$, i.e.,
\beq
\gamma_Y(x,x') = \frac {N_1}{n_Y} \int  \Psi_N(x,x_2,\dots,x_{N_1},Y) \Psi_N(x',x_2,\dots,x_{N_1},Y)^*\,d^3\!x_2\cdots d^3\!x_{N_1}.
\eeq
Note that $0\leq \gamma_Y\leq 1$ and $\Tr \gamma_Y = N_1$. Moreover,
$\int n_Y\,dY = 1$ and $\int n_Y \gamma_Y \, dY= \gamma_1$, the
one-particle reduced density matrix for the $X$-particles.  

Let $P$ be a projection operator, and let $\gamma$ denote any fermionic density matrix, which is an operator that satisfies $0\leq \gamma\leq 1$ and $\Tr \gamma = N_1$. Let $W_\pm$ be two bounded positive semi-definite operators, and let $W=W_+-W_-$.
 For any $\delta>0$, we have
\begin{eqnarray*}
\Tr[ \gamma W] &=& \Tr[PW] + \Tr[ (\gamma-1)PWP] \\ &&+ \Tr\left[ \gamma \left( (1-P) W P + PW(1-P)+ (1-P) W (1-P)\right)\right] \\ &\geq& \Tr[PW_+](1-\delta) - \Tr[PW_-](1+\delta) \\ && -  \left(1+\delta^{-1}\right) \left( \|W_+\|+\|W_-\|\right)  \Tr[\gamma(1-P)] - \|W\| \Tr[P(1-\gamma)],
\end{eqnarray*}
with $\|\, \cdot\, \|$ denoting operator norm. 
Now let $P\equiv P_{N_1}$ be the operator defined in (\ref{perker}), and $W=W_Y$. We choose $W_+$ to be the terms in (\ref{defwy}) containing $U(x)$, and $W_-$ the ones containing $w_R(x)$. We then have, using $\int U(x)\, d^3\!x = 4\pi$,  
\begin{eqnarray*}
\Tr[P W_+] &=& \frac {\Tr[P_{N_1}]}{L^3}  \sum_{\{ j\, : \, y_j \in \widetilde Y\}}  (1-\eps) a \int_{[0,L]^3}U(x-y_j)\, d^3\!x \\ &\geq& \frac {\Tr[P_{N_1}]}{L^3}  (1-\eps)4\pi a  \left[N_2-I_R(Y)- \const \frac {L^2}{R^2} \right].
\end{eqnarray*}
The last term in square brackets bounds the number of $y_j$'s in $\widetilde Y$ that are at least a distance $R$ away from the boundary of the box. Since the distance between the $y_j$'s is bigger than $2R$ by assumption, the number of such $y_j$'s close to the boundary is bounded by $\const L^2/R^2$. By Lemma~\ref{lemir}, 
\beq\label{iy}
\int n_Y I_R(Y)\, dY = \left\langle \Psi_N\left| I_R(Y) \right| \Psi_N\right\rangle \leq \const N (R^3\rho)^{2/3}
\eeq
if $\Psi_N(X,Y)$ is an approximate ground state.  As already noted in Eq. (\ref{trpm}), $\Tr[P_{N_1}]$ can be replaced by $N_1$ in the thermodynamic limit. 

Analogously, using (\ref{intw}),  we get an upper bound
$$
\Tr[P W_-] \leq \const \frac{a R^2}{\eps s^2} \frac {N_2 \Tr[P_{N_1}]}{L^3} .
$$
Moreover, using (\ref{normw}) and the fact that the distance between $y_j$'s contributing to $W_Y$ is at least $2R$, we find that
$$
\|W_Y\|_\infty \leq \|W_+\|+\|W_-\|\leq  \left( \frac {3a}{R^3-R_0^3} + \const \frac {a}{\eps s^2 R} \right).
$$
The {\it a priori} bound in Lemma~\ref{lemtr} implies that, for large enough $N$,  
\beq
\int n_Y \Tr[\gamma_Y (1-P)]\, dY = \Tr[\gamma_1 (1-P)] \leq C N (a^3\rho)^{1/6},
\eeq
where $\gamma_1$ is the one-particle density matrix (for the $X$-particles) of any approximate ground state. The same bound is true for $\Tr[P(1-\gamma_1)]= \Tr[\gamma_1(1-P)]+ \Tr[P-\gamma_1]$, since $N_1^{-1} \Tr[P-\gamma_1]\to 0$ as $N_1\to \infty$ (see (\ref{trpm})). Hence, collecting all the bounds, and applying the same arguments also to the second term in (\ref{hamxy}), we arrive at the lower bound
\begin{eqnarray*}
 \lim_{L\to\infty} \frac 1{L^3}  E_0(N_1,N_2,L) &\geq&   \frac 35 \left(6\pi^2\right)^{2/3} \big[\rho_1^{5/3}+\rho_2^{5/3}\big] \\ &&  + 8\pi a \rho_1 \rho_2 \left( 1 -\eps -\delta - s^2 (6\pi^2 \rho)^{2/3}- C\frac{R^2}{\eps s^2}  \right) -C a\rho^2   (R^3\rho)^{2/3} \\ &&  - C \rho (a^3\rho)^{1/6}\left(1+\frac 1\delta\right) \left( \frac {a }{R^3-R_0^3} +  \frac {a }{\eps s^2 R}\right) 
\end{eqnarray*}
for some $C>0$. 

We choose 
\beq\label{chos}
R=\rho^{-1/3}\big(a\rho^{1/3}\big)^{3/26} \ , \ s=\rho^{-1/3}\big(a\rho^{1/3}\big)^{1/26}\ , \  \eps=\delta=\big(a\rho^{1/3}\big)^{1/13}
\eeq
and obtain, for small $\rho$,
$$
\lim_{L\to\infty} \frac 1{L^3} E_0(N_1,N_2,L) \geq  \frac 35 \left(6\pi^2\right)^{2/3} \big[ \rho_1^{5/3}+\rho_2^{5/3}\big]  + 8\pi a \rho_1\rho_2 -  \const a\rho^2  \big(a\rho^{1/3}\big)^{1/13}.
$$
This finishes the proof of the lower bound.

\section{The Two-Dimensional Gas}\label{2dsect}

We now comment on the necessary changes in considering the 2D gas instead of the 3D gas. 
We start with the lower bound to the ground state energy. The analogue of Lemma~\ref{dyson} in 2D is the following lemma, which generalizes the corresponding result used for bosons in 2D in \cite{LY2001}. Its proof can again be found in the appendix. 

\begin{lem} \label{dyson2d}
For $R>R_0$, let $\theta_R(x)$ denote the characteristic function 
of a disc of radius $R$ centered at the origin, i.e., $\theta_R (x) =1 $ if 
$|x| <R$ and $=0$ otherwise. Let $\chi(p)$ be a radial function, 
$0\leq \chi (p) \leq 1$, such that $h(x)\equiv \widehat{(1-\chi)}(x)$ 
is bounded and integrable. Let 
\beq\label{deffr2d} 
f_R(x)= \sup_{|y|\leq R} | h(x-y) - h(x) |, 
\eeq
  and 
\beq\label{defwr2d} w_R(x)= \frac{2}{\pi} f_R(x) \int_{\R^2}
  f_R(y)\,d^2\!y.  
\eeq 
Let $U(x)$ be any positive, radial function, supported in the annulus $R_0\leq |x|\leq R$, with 
\beq\label{condu}
\int_{\R^2} U(x) \ln(|x|/a) \,d^2\!x = 2\pi.
\eeq
Then, for any $\eps>0$,
\beq\label{topr2d}
-\nabla \chi(p) \theta_R(x) \chi(p) \nabla + \half v(x) 
\geq (1-\eps) U(x) - \frac 1\eps \left[(2\pi)^{-1} \mbox{$\int U(y) \,d^2\!y$} \right] w_R(x).
\eeq
\end{lem}

In the application, we choose, as in \cite{LY2001},
\beq
U(x) = \left\{ \begin{array}{ll}  \nu(R)^{-1} & {\rm for\ } R_0\leq |x|\leq R \\ 0 & {\rm otherwise}, \end{array} \right.
\eeq
with $\nu(R)$ determined by condition (\ref{condu}), i.e.,
\beq
\nu(R) = \int_{R_0}^R \ln (r/a) r \,dr = \frac 14 \left[ R^2 \ln\frac{R^2}{a^2 e} - R_0^2 \ln \frac{R_0^2}{a^2 e}\right].
\eeq
Using that $a\leq R_0\leq R$ we get the bounds
\beq
\half (R^2-R_0)^2 \left( \ln R/a - \half\right) \leq \nu(R) \leq \half R^2 \ln R/a,
\eeq
from which, in turn, we get upper and lower bounds on $\int U(x) \,d^2\! x= \nu(R)^{-1} \pi (R^2-R_0^2)$. 

Moreover, we again choose $\chi(p)$ as in (\ref{defchi}), with $R\leq \const s$. Inequalities (\ref{intw})--(\ref{normw}) then have to be replaced in the 2D case by 
\beq\label{intw2d} 
|w_R(x)| \leq \const \frac{R^2}{s^4} \ \quad {\rm and\quad } \int |w_R(x)|\, d^2\!x \leq \const \frac {R^2}{s^2} 
\eeq 
and 
\beq\label{normw2d} 
\sum_{i=1}^N w_R(x-y_i) \leq \const \frac 1{s^2} 
\eeq 
in case that $|y_i-y_j|\geq 2R$ for all $i\neq j$. 

The {\it a priori} bounds of Subsect.~\ref{apss} can be obtained also in the 2D case. The proof of Lemma~\ref{lemtr} works in the same way, with the appropriate changes in the expression of the kinetic and interaction energy, of course. For the proof of Lemma~\ref{lemir}, we note that the analogue of the inequality (\ref{lyineq}) does {\it not} hold in 2 dimensions. However, a \lq relativistic\rq\ version of it is true, namely that 
\beq\label{lyineq2d}
\sum_{i=1}^{N_2} \frac 1{\delta_i} \leq \const \sum_{i=1}^{N_2} \sqrt{ -\Delta_i}
\eeq
on antisymmetric functions of $N_2$ variables $y_i\in\R^2$. Ineq. (\ref{lyineq2d}) can be proved in a similar way as the proof of (\ref{lyineq}) in \cite{LYau}. It implies that 
\beq\label{ineqI2d}
\langle \Psi_N| I_R(y_1,\dots,y_{N_2}) |
\Psi_N\rangle \leq 2R \,\Tr [ \sqrt{-\Delta}\, \gamma] \leq 2R \left( \Tr [ -\Delta\, \gamma] \right)^{1/2} \left( \Tr \gamma\right)^{1/2} \leq  \const N (R^2\rho)^{1/2}
\eeq
in 2D, replacing (\ref{ineqI}). Here $\gamma$ denotes the one-particle density matrix (for the $Y$ particles) of $\Psi_N(X,Y)$, and  we have used Schwarz's inequality as well as the assumption $\Tr [-\Delta\, \gamma] \leq  \const N \rho$ for an approximate ground state.
 
With the {\it a priori} bounds in hand, we can proceed along the same lines as in Subsect.~\ref{ssput} to obtain a lower bound to the ground state energy. The optimal choice of the free parameters $\eps$, $\delta$, $R$ and $s$ in 2D turns out to be 
$$
R=\rho^{-1/2}\frac 1{|\ln(a^2\rho)|^{3/20}} \ , \ s=\rho^{-1/2}\frac 1{|\ln(a^2\rho)|^{1/20}}
\ , \  \eps=\delta=\frac 1{|\ln(a^2\rho)|^{1/10}}.
$$
This yields the lower bound in Theorem~\ref{T2}.

Our last task is to derive the upper bound in Theorem~\ref{T2}. It turns out that obtaining this bound is actually much easier than in the 3D case. The reason for the rather complicated construction in 3D was the very small interaction energy $\sim a\rho$ per particle, which forced us to choose the particle number in a box to be quite large, namely $n\gg 1/(a^3\rho)$, in order to have negligible finite size effects. This resulted in a trial wave function with very small norm. In 2D, however, it is possible to choose the particle number in each box much smaller, such that the norm of the trial wave function is close to one. If we take the analogous function as in (\ref{tri1})--(\ref{tri2}), with $s=2R$ and with $\varphi(x)$ now being the solution to the zero-energy scattering equation in 2D, cut off at an appropriate radius $R$, then a simple bound as in the proof of Lemma~\ref{lemg} shows that
\beq
\langle\psi|\psi\rangle \geq 1 - \const n (R^2\rho) .
\eeq
Hence we have to choose the box size $\ell$ and $R$ such that $n R^2 \rho^2\ll 1$, with $n\sim\rho \ell^2$. Moreover, the restriction on having a negligible finite size effect is $n^{1/2} \gg |\ln(a^2\rho)|$ (compare with (\ref{estiI})). If we choose $R=\rho^{-1/2} |\ln(a^2\rho)|^{-\alpha}$ for large enough $\alpha$, then all these conditions are easily fulfilled. In calculating the kinetic energy in (\ref{defII}) and (\ref{defIII}), we can then just use the simple bounds $g(x)\leq 1$ and $f(x)\leq 1$. We demonstrate this on the analogue of the term II in (\ref{defII}) in 2D. Namely, with $\xi(x)$ given as in (\ref{defxi}),
\begin{eqnarray}\nonumber
&& \int \left[ |\nabla_X F(X,Y)|^2  + \half v_{XY} F(X,Y)^2\right] D_n(X)^2 D_m(Y)^2 G_n(X)^2 G_m(Y)^2 \,dX\,dY \\ && \leq \sum_{i=1}^n \sum_{j=1}^m \int \xi(x_i-y_j) D_n(X) D_m(Y) \, dX\,dY  = \int \rho^{\rm D}_n(x) \rho^{\rm D}_m (y) \xi(x-y) \,d^3\!x\,d^3\!y. 
\end{eqnarray}
Here we have also used the fact that the integrand vanishes whenever two particles of the same kind are closer together then a distance $s\geq 2R$, in order for (\ref{holds}) to hold. We can then proceed using Young's inequality on the last term, as in (\ref{inte})--(\ref{inte3}). 
The leading term from the interaction energy then comes from 
\beq\label{er2d}
\int_{|x|\leq R} (|\nabla \varphi(x)|^2 + \half v(x)|\varphi(x)|^2) \,d^2\!x =\frac{ 2\pi}{\ln(R/a)} \leq \frac {4\pi }{|\ln(a^2\rho)|} \left( 1 + \const \frac{\ln|\ln(a^2\rho)|}{|\ln(a^2\rho)|} \right). 
\eeq 
The other terms in the upper bound can be treated in the same way. It turns out that, choosing $\alpha$ large enough, all the other error terms besides the one in (\ref{er2d}) are of lower order in $a^2\rho$. We omit the details. This results in the upper bound stated in Theorem~\ref{T2}.

\appendix
\section{Proof of Lemmas~\ref{dyson} and~\ref{dyson2d}}

We start with the three-dimensional case, Lemma~\ref{dyson}. 
It suffices to show that the operator inequality (\ref{topr}) holds for the expectation value with any smooth function $\psi(x)$ of compact support. Given such
  a $\psi(x)$, define the function $\xi(x)$ by its Fourier transform
  $\widehat \xi(p) = \chi(p) \widehat \psi(p)$. We thus have to
  show that
 \beq\label{toshow}
\int_{|x|\leq R} \left[ |\nabla
    \xi(x)|^2 + \half v(x) |\psi(x)|^2\right] d^3\!x \geq
  \int_{\R^3} \left[ (1-\eps) a U(x) |\psi(x)|^2 - \frac a\eps w_R(x)
    |\psi(x)|^2\right] \,d^3\!x.  
\eeq

Let $\varphi(x)$ denote the solution to the zero-energy scattering
equation (\ref{scatteq}), subject to the boundary condition $\lim_{|x|\to \infty}
\varphi(x)=1$.  Let $\nu$ be a complex-valued function on
the unit sphere $\Ss^2$ , with $\int_{\Ss^2} |\nu|^2 = 1$.  We use the
same symbol for the function on $\R^3$ taking values $\nu(x/|x|)$.  For $\psi(x)$ as above, consider the expression
$$
A\equiv \int_{|x|\leq R} \nu(x) \nabla\xi^*(x)\cdot \nabla \varphi(x) \, d^3\!x + \half \int v(x) \psi(x)^* \varphi(x) \nu(x) \,d^3\!x.
$$ 
We note that the last integral makes sense even in the case when $v(x)$ has a hard core; in this case, $\half v(x)\varphi(x)$ has to be interpreted as the (non-negative) measure $\Delta \varphi(x)$ (see Eq.~(\ref{scatteq})). By using the Cauchy-Schwarz inequality, we can obtain the upper bound
\begin{eqnarray*}
|A|^2 &\leq& \left(\int_{|x|\leq R} \left[ |\nabla\xi(x)|^2 + \half v(x) |\psi(x)|^2\right]\, d^3\!x\right)  \left(\int_{|x|\leq R} \left[ |\nabla \varphi(x)|^2 + \half v(x) |\varphi(x)|^2\right] \nu(x)^2 \,d^3\!x\right).
\end{eqnarray*}
Since $\varphi(x)$ is a radial function, the angular integration in the last term can be performed by using $\int_{\Ss^2} |\nu|^2 = 1$. The remaining expression is then bounded by $a$ because of $\int_{\R^3} \left( |\n \varphi(x)|^2 +
\half v(x) |\varphi(x)|^2 \right)\,d^3\!x = 4\pi a$, as pointed out in the beginning of Section~\ref{upsect}.  Hence we arrive at 
\beq\label{comb}
\int_{|x|\leq R} \left[ |\nabla\xi(x)|^2 + \half v(x) |\psi(x)|^2\right] \,d^3\!x \geq \frac {|A|^2}a ,
\eeq
for any choice of $\nu$ as above. It remains to derive a lower bound on $|A|^2$. 

Note that $\varphi(x)$ is a radial function with $|\nabla\varphi(x)|=a/R^2$ for $|x|=R$. Hence we obtain, by partial integration,
$$
\int_{|x|\leq R} \nu(x) \nabla\xi^*(x)\cdot \nabla \varphi(x)  \,d^3\!x = - \int_{|x|\leq R} \xi^*(x) \nu(x) \Delta \varphi(x) \,d^3\!x + \frac a{R^2} \int_{|x|=R}
\xi^*(x) \nu(x) \, d\omega_R ,
$$
where $d\omega_R$ denotes the surface measure of the ball of radius $R$, and we used the fact that $\nabla \nu(x)\cdot \nabla \varphi(x)=0$. Now, by definition of $h(x)$, $\xi(x)= \psi(x) - (2\pi)^{-3/2} h * \psi(x)$, where $*$ denotes convolution, i.e., $h*\psi(x) = \int h(x-y)\psi(y) \,d^3\!y$.  Using the zero-energy scattering equation (\ref{scatteq}) for $\varphi(x)$, we thus see that  
\begin{eqnarray}\nonumber
A =  \frac a{R^2} \int_{|x|=R} \psi^*(x) \nu(x) \,d\omega_R &-& (2\pi)^{-3/2} \frac a{R^2} \int_{|x|=R} (h*\psi)^*(x) \nu(x) \,d\omega_R
 \\ &+& (2\pi)^{-3/2} \int_{|x|\leq R}( h*\psi)^*(x) \nu(x) \Delta \varphi(x)\,d^3\!x.  \label{of}
\end{eqnarray} 
The last two terms on the right side of (\ref{of}) can be written as (note that $h$ is a real-valued function)
\beq\label{expr}
(2\pi)^{-3/2} \int \psi^*(x) \left[ \int  h(y-x) \,d\mu(y)\right] \,d^3\!x ,
\eeq
where $d\mu$ is a (non-positive) measure supported in the ball of
radius $R$. Explicitly, $d\mu(y) = - a R^{-2}
\nu(y) \delta(|y|-R)d^3\!y + \nu(y) \Delta \varphi(y)d^3\!y $.
Note that $\int d\mu(y) = 0$, and also $\int d|\mu(y)| = 2
a \int_{\Ss^2} |\nu| \leq 2 a \sqrt{4\pi} $ (by Schwarz's inequality).  Hence
$$
 \left| \int  h(y-x) \,d\mu(y)\right| \leq 2 a \sqrt{4\pi} f_R(x),
$$
with $f_R(x)$ defined in (\ref{deffr}). The expression (\ref{expr}) is thus bounded from below by 
\beq\label{15}
(\ref{expr}) \geq - (2\pi)^{-3/2} 2a \sqrt{4\pi} \int |\psi(x)| f_R(x)\,d^3\!x \geq - a \left( \int |\psi(x)|^2 w_R(x) \,d^3\!x \right)^{1/2}, 
\eeq
where we used Schwarz's inequality as well as the definition of $w_R(x)$ (\ref{defwr}) in the last step. Note that this last expression is independent of $\nu(x)$.

The only place where $\nu(x)$ still enters is the first term on the right side of (\ref{of}). By choosing $\nu(x)$ to be the
restriction of $\psi(x)$ to the sphere of radius $R$, appropriately
normalized, we obtain from (\ref{of})--(\ref{15}) 
$$
A \geq \frac aR \left( \int_{|x|=R} |\psi(x)|^2 \,d\omega_R\right)^{1/2}  
-  a \left( \int |\psi(x)|^2 w_R(x) \,d^3\!x \right)^{1/2}. 
$$
Using again the Cauchy-Schwarz inequality, we see that, for any $\eps>0$, 
\beq\label{comb2}
|A|^2 \geq \frac{a^2}{R^2} (1-\eps)\int_{|x|=R} |\psi(x)|^2 \,d\omega_R - \frac {a^2}\eps  
 \int |\psi(x)|^2 w_R(x)\, d^3\!x.
\eeq
 In combination with (\ref{comb}) this proves the desired result
 (\ref{toshow}) in the special case when $U(x)$ is a radial
 $\delta$-function sitting at a radius $R$, i.e., $U(x)= R^{-2}
 \delta(|x|-R)$. The case of a general potential $U(x)$ follows simply by integrating this
 result (i.e., Ineq. (\ref{toshow}) for this special $U(x)$) against
 $u(R) R^2 dR$, with $u(R)=U(x)$ for $|x|=R$, noting that $\int u(R) R^2 dR=1$ and that $w_R(x)$ is
 pointwise monotone increasing in $R$.

The proof in the two-dimensional case, Lemma~\ref{dyson2d}, follows exactly the same lines. Note, however, that the solution to the zero-energy scattering equation can not be normalized by $\lim_{|x|\to\infty}\varphi(x)=1$ in 2D, but we can normalize it such that $\varphi(x)=1$ for $|x|=R$. It then follows that $\varphi(x)=\ln(|x|/a)/ \ln(R/a)$ for $R_0\leq |x|\leq R$, and also that $\int_{|x|\leq R} (|\nabla \varphi(x)|^2 + \half v(x)|\varphi(x)|^2) \,d^2\!x = 2\pi / \ln(R/a)$ and $\int \Delta\varphi(x) \,d^2\!x =2\pi /\ln(R/a)$ (see the appendix in \cite{LY2001}). The rest of the proof is unchanged, with the result that
\beq
\int_{|x|\leq R} \left[ |\nabla\xi(x)|^2 + \half v(x) |\psi(x)|^2\right] \,d^2\!x \geq \frac{1}{\ln(R/a)} \left[ (1-\eps) \frac 1R \int_{|x|=R} |\psi(x)|^2 \,d\omega_R - \frac {1}\eps  
 \int |\psi(x)|^2 w_R(x) \,d^2\!x \right]  
\eeq
instead of (\ref{comb}) and (\ref{comb2}). Multiplying this inequality with $u(R) R \ln(R/a)$, where $u(R)=U(x)$ for $|x|=R$, and integrating over $R$ using (\ref{condu}), we arrive at the desired result. 

\begin{acknowledgments}
The authors are grateful to Jakob Yngvason for several helpful discussions
and remarks.
The work was supported in part by the NSF grants
PHY 0139984-A01 (EHL), PHY 0353181 (RS) and DMS-0111298 (JPS); 
by an A.~P.~Sloan Fellowship (RS); by
EU grant HPRN-CT-2002-00277 (JPS),
by MaPhySto -- A Network in Mathematical Physics and
Stochastics funded by The Danish National Research Foundation (JPS),
and by grants from the Danish research council (JPS).
\end{acknowledgments}


\begin{thebibliography}{99}

\bibitem{huang}
K.~Huang, C.N.~Yang, {\it Quantum-Mechanical Many-Body Problem with Hard-Sphere Interaction}, Phys. Rev. {\bf 105}, 767--775 (1957).

\bibitem{lee}
T.D.\ Lee, C.N.\ Yang, {\it Many-Body Problem in Quantum Mechanics and Quantum Statistical Mechanics}, Phys. Rev. {\bf 105}, 1119--1120 (1957). 

\bibitem{fetter}
A.L.\ Fetter, J.D.\ Walecka, {\it Quantum Theory of Many-Particle Systems}, McGraw--Hill, New York (1971). 

\bibitem{Lenz}
W.~Lenz, {\it Die Wellenfunktion und Geschwindig\-keits\-verteilung des entarteten Gases}, Z. Phys. {\bf 56}, 778--789 (1929).

\bibitem{LY1998}
E.H.\ Lieb, J.\ Yngvason, {\it Ground State Energy of the Low
Density Bose Gas}, Phys. Rev. Lett. \textbf{80}, 2504--2507 (1998).

\bibitem{dyson}
F.J.\ Dyson, {\it Ground-State Energy of a Hard-Sphere Gas}, Phys. Rev. {\bf 106}, 20--26 (1957).

\bibitem{schick} M.~Schick, {\it Two-dimensional System of Hard Core Bosons}, 
Phys. Rev. A {\bf 3}, 1067--1073 (1971). 

\bibitem{hfm} D.~F.~Hines, N.~E.~Frankel, D.~J.~Mitchell,
{\it Hard disc Bose gas}, Phys.~Lett. {\bf 68A}, 12--14 (1978).

\bibitem{LY2001}
E.H.\ Lieb, J.\ Yngvason, {\it The Ground State Energy of a Dilute Two-Dimensional Bose Gas}, J. Stat. Phys. \textbf{103}, 509--526 (2001).  

\bibitem{baker} G.A.~Baker, {\it Singularity Structure of the Perturbation Series
for the Ground-State Energy of a Many-Fermion System},
Rev. Mod. Phys. {\bf 43}, 479--531 (1971). 

\bibitem{hammer}
H.-W.\ Hammer, R.J.\ Furnstahl, {\it Effective field theory for dilute Fermi systems}, Nucl. Phys. A {\bf 678}, 277--294 (2000).

\bibitem{RSfermiT}
R. Seiringer, {\it The Thermodynamic Pressure of a Dilute Fermi Gas},
in preparation.

\bibitem{ruelle}
D.\ Ruelle, {\it Statistical Mechanics. Rigorous Results}, World Scientific (1999).

\bibitem{robinson}
D.W.\ Robinson, {\it The Thermodynamic Pressure in Quantum Statistical Mechanics}, Springer Lecture Notes in Physics, Vol.~9 (1971).

\bibitem{anal} E.~H. Lieb, M.~Loss, {\it Analysis}, Amer. Math. Soc.
(2001).

\bibitem{GS} G.M.\ Graf, J.P.\ Solovej, \textit{A correlation
estimate
with applications to quantum systems with Coulomb interactions},
Rev. Math. Phys. {\bf 6}, 977--997 (1994).

\bibitem{LYau}
E.H.\ Lieb, H.-T.\ Yau, {\it The Stability and Instability of Relativistic Matter}, Commun. Math. Phys. {\bf 118}, 177--213 (1988). 

\bibitem{liyau}
P.\ Li, S.-T.\ Yau, {\it On the Schr\"odinger equation and the eigenvalue problem}, Commun. Math. Phys. {\bf 88}, 309--318 (1983). 


\end{thebibliography}
\end{document}